\documentclass[ 10pt, conference]{IEEEtran}
\IEEEoverridecommandlockouts
\usepackage{lipsum}
\usepackage{bbm}
\usepackage{lettrine}
\usepackage{csquotes}
\usepackage{amsmath,amssymb,amsfonts}

\usepackage[ruled,linesnumbered]{algorithm2e}
\usepackage{array}
\usepackage[caption=false,font=normalsize,labelfont=sf,textfont=sf]{subfig} 
\usepackage{textcomp}
\usepackage[normalem]{ulem}   
\usepackage{stfloats}
\usepackage{url}
\usepackage{verbatim}
\usepackage{graphicx}
\usepackage[dvipsnames]{xcolor}
\usepackage{soul}             
\usepackage{enumitem}
\usepackage{arydshln}

\usepackage{pgfplots}
\usetikzlibrary{patterns}
\pgfplotsset{compat=newest}

\usepackage[short]{optidef}
\usepackage{cite}

\usepackage{booktabs} 
\usepackage[nolist,nohyperlinks]{acronym}

\def\BibTeX{{\rm B\kern-.05em{\sc i\kern-.025em b}\kern-.08em
    T\kern-.1667em\lower.7ex\hbox{E}\kern-.125emX}}

\DeclareMathOperator*{\argmax}{\arg\!\max}

\begin{document}

\title{Optimal SSB Beam Planning and UAV Cell Selection for 5G Connectivity on Aerial Highways}

\author{
\vspace{-0.3cm}
\IEEEauthorblockN{Matteo Bernabè\IEEEauthorrefmark{1}\IEEEauthorrefmark{2}, 
David López-Pérez\IEEEauthorrefmark{3},
Nicola Piovesan\IEEEauthorrefmark{1},
Giovanni Geraci\IEEEauthorrefmark{4}\IEEEauthorrefmark{5}, and
David Gesbert\IEEEauthorrefmark{2}}
\\ \vspace{-0.1cm}
\normalsize\IEEEauthorblockA{\emph{\IEEEauthorrefmark{1}Huawei Technologies, France \quad \IEEEauthorrefmark{2}EURECOM, France \quad \IEEEauthorrefmark{3}iTeam, Universitat Politècnica de València, Spain } 
}
\normalsize\IEEEauthorblockA{\emph{
\IEEEauthorrefmark{4}Telef\'{o}nica Research, Spain \quad \IEEEauthorrefmark{5}Universitat Pompeu Fabra, Spain \quad }
}
\thanks{
This research was supported by 
\emph{a)} the Generalitat Valenciana, Spain, through the  CIDEGENT PlaGenT,  Grant CIDEXG/2022/17,  Project iTENTE, 
\emph{b)} the action CNS2023-144333, financed by MCIN/AEI/10.13039/501100011033 and the European Union “NextGenerationEU”/PRTR,
\emph{c)} the Spanish State Research Agency through grants PID2021-123999OB-I00, CEX2021-001195-M, and CNS2023-145384, 
\emph{d)} the UPF-Fractus Chair on Tech Transfer and 6G, and 
\emph{e)} the Spanish Ministry of Economic Affairs and Digital Transformation.
} 
\vspace{-0.9cm}
}

\bstctlcite{IEEEexample:BSTcontrol}

\maketitle

\begin{abstract}\label{sec:Abstract}
In this article, 
we introduce a method to optimize 5G \ac{mMIMO} connectivity for \acp{UAV} on aerial highways through 
strategic cell association.
\acp{UAV} operating in 3D space encounter distinct channel conditions compared to traditional \ac{gUE}; 
under the typical \ac{LoS} condition, \acp{UAV} perceive strong \ac{RSRP} from multiple cells within the network, resulting in a large set of suitable serving cell candidates and in low \ac{SINR} due to high interference levels.
Additionally, a downside of aerial highways is to pack possibly many \acp{UAV} along a small portion of space which, when taking into account typical \ac{LoS} propagation conditions, results in high channel correlation and severely limits spatial multiplexing capabilities. 
In this paper, we propose a solution to both problems based on the suitable selection of serving cells based on a new metric which differs from the classical terrestrial approaches based on maximum \ac{RSRP}.
We then introduce an algorithm for optimal planning of \ac{SSB} beams for this set of cells, 
ensuring maximum coverage and effective management of \acp{UAV} cell associations. 
Simulation results demonstrate that our approach significantly improves the rates of \acp{UAV} on aerial highways, up to four times in achievable data rates, without impacting ground user performance.

\end{abstract}

\acresetall

\section{Introduction}\label{sec:Intoduction}

\Acp{UAV} have emerged as a key technology across multiple market sectors,
including photography, infrastructure inspection and disaster management~\cite{zeng2020uav,Wu2021,8470897}.
Only in recent years have \acp{UAV} become an integral part of the urban scenario as well \cite{KanMezLoz2021}. 
Overall, \ac{UAM}, 
including future transportation, cargo drones, and other civil applications, 
is expected to play a disruptive role in future markets, 
with recent reports projecting its value to reach 5.1 billion U.S. dollars by 2028~\cite{MarketGrowthReport}.
However, the burgeoning interest in UAVs within urban scenarios raises critical challenges:
\textit{i)} development of regulation for secure management in urban skies, 
\textit{ii)} supporting reliable connectivity in the sky enabling \ac{BVLoS} applications. 
In terms of regulation, 
industries and regulatory bodies are working towards creating a highway system for the sky, 
namely \acp{AH};
similar to traditional ground road scenarios, 
\acp{AH} ---also often denoted as UAV corridors--- are defined trajectories that \acp{UAV} must follow while pursuing their tasks~\cite{PilotsHandbook_FAA, 9566514}.

In recent years, 
studies identified in cellular networks the key technologies to enable \ac{BVLoS} services \cite{9768113,9681624,geraci2022integrating, 9839125, DoPulFod2024}.
Nevertheless, 
the optimal integration of cellular-connected \acp{UAV} in terrestrial networks remains a challenge, especially when considering \acp{AH}. 
Few pioneering works in the literature addressed this problem. 
In the context of \ac{AH} supported by 4G networks, 
authors in~\cite{10021675} considered a set of uptilted sectors to serve the \ac{AH} while providing, under specific assumptions, an analytical framework for outage probability. 
Similarly,
authors in~\cite{chowdhury2021ensuring} deployed a new set of uptilted antennas 
while proposing a solution to mitigate the generated interference to the ground.
In our previous work~\cite{10001469}, we introduced an ADAM-based solution to optimize the vertical tilt of 4G base stations for \ac{UE} on the ground and along \acp{AH} without the need for new infrastructure. 
Driven by similar motivations, 
authors in~\cite{karimibidhendi2023optimizing} and in~\cite{benzaghta2023designing} respectively proposed quantization theory- and Bayesian optimization-based approaches to design cell antenna tilt and transmit power and optimally cover both \acp{gUE} and \acp{UAV} within \acp{AH}.

In 5G, \ac{mMIMO} offers a paradigm shift and is capable of enhancing \acp{UAV} communications too \cite{8528463,8869706, 9374639}. 
Previous studies that showed significant advantages of optimizing serving cells in \ac{mMIMO} \acp{UDN} \cite{7314981, 7247514}, suggest that similar principles benefit \ac{UAV} in \ac{UMa} scenarios. Indeed, the typical \ac{LoS} for \acp{UAV} creates similar cell association dynamics as in \acp{UDN}, therefore controlling cell association along the \ac{AH} become crucial for improving \acp{UAV} connectivity. Unlike the real-time centralized schedulers proposed in previous work, new solutions are needed to tackle the problem at the radio access network planning stage.

In this work
we demonstrate how
leveraging the prior knowledge of the \ac{AH} trajectory to plan and control
the transmitted \ac{SSB} beams, 
and in turn the \acp{UAV} cell association processes, 
allows to efficiently optimize connectivity on \acp{AH}.
Specifically,  
we propose a new metric to optimally define the set of cells aimed to serve \acp{UAV}, 
by jointly considering multiplexing capability, average channel gain, and interference of each cell.
Furthermore, we propose an \ac{eGA} to optimally select \ac{SSB} beams and their transmit power within the set of identified cells, 
thereby ensuring desired cell association.\footnote{Extensions of this work can be found in~\cite{10569086}.}

\section{System Model}\label{sec:SystemModel}

We focus on a downlink, interference-limited scenario, 
with models as defined by the \ac{3GPP}. 

\subsubsection*{Network deployment}

We consider a cellular network operating in a sub-6~GHz band (FR1), 
with carrier frequency  $f_c$ and bandwidth $B_0$.
The network layout consists of 19 sites, 
organized in a 2-tier hexagonal grid,
with an inter-site distance $d_{\rm ISD}$. 
Each site is composed of three sectors\footnote{
In the rest of the paper, the terms \enquote{sector} and \enquote{cell} are used as synonyms.}, 
each covering $120^\circ$. 
The complete set of sectors is denoted by $\mathcal{B}$, 
with $N_{\rm BS}$ denoting its cardinality.
Full frequency reuse is applied in all sectors. 
Each sector contains a \ac{UPA} antenna panel located at a height $h_{\rm BS}$, 
consisting of $M$ single vertically polarized antenna elements, 
arranged in $M_h$ horizontal and $M_v$ vertical rows.
The total number of \acp{PRB} available is $N_{\rm PRB}$, 
each with a bandwidth of $B_{\rm PRB}$. 

\subsubsection*{Terrestrial users}

Assuming a fully loaded scenario, we consider a total of $N_g$ \acp{gUE} randomly distributed in all cells.
Additionally, 
to capture the dynamic nature of the network, 
the positions of \acp{gUE} randomly vary over time.

\subsubsection*{Aerial users}

An \ac{AH} ${\bf r}_{\rm AH}$, spanning a total length of $L_{\rm AH}$, is positioned over multiple cell centres and edges of our scenario at an altitude of $h_{\rm AH}$.
For simplicity, 
we consider that the \ac{AH} is divided into $N_{\rm seg}$ consecutive segments.
Over the aforementioned \ac{AH}, 
a total number of $N_{a}$ \acp{UAV} are evenly spaced with constant \ac{IUD} $d_{\rm IUD}$; 
all the defined \acp{UAV} move along the \ac{AH} while maintaining same $d_{\rm IUD}$.
To maintain continuous aerial traffic, 
note that when one \ac{UAV} exits, another enters the \ac{AH}. 

We denote by $\mathcal{G}$ the set of all \acp{gUE}, by $\mathcal{A}$ the set of \acp{UAV} and by $\mathcal{U}$ the set of all \acp{UE}, such that $\mathcal{U} = \mathcal{G} \cup \mathcal{A}$.

\subsection{Channel Model}

We consider the \ac{3GPP} statistical channel models defined in \cite{3GPP38901} and~\cite{3GPP36777}.

\subsubsection*{Large-scale fading}

For each \ac{UE} $u \in \mathcal{U}$ and sector $b \in \mathcal{B}$, the large-scale channel between them is obtained from the \ac{LoS} probability $P_{\rm LoS}$, 
path loss gain $\rho_{u_b}$,
antenna element gain $g_{u,b}$, 
and shadow fading gain $\tau_{u_b}$.
Note that the shadow fading gain is modelled as spatially correlated as per the \ac{3GPP} recommendations. 
Using the 3GPP models,
we can then define the large-scale gain $\beta_{u,s}$ as follows:
\begin{equation} \label{eq:LargeScaleGain}
    \beta_{u,b} = \rho_{u,b} \; \tau_{u,b} \; g_{u,b}.
\end{equation}

\subsubsection*{Small-scale fading}

To model the small-scale fading, 
the downlink complex channel vector between each \ac{UE} $u$ and each antenna element $m$ of each sector $b$ is defined as follows:
\begin{equation}\label{eq:ComplexChannelRician}
    {\bf h}^{\rm dl}_{u,b} =
    \sqrt{\frac{K}{1+K}} \, {\bf h}^{\rm LOS}_{u,b}
    +
    \sqrt{\frac{1}{1+K}} \,  {\bf h}^{\rm NLOS}_{u,b} ,
\end{equation}
where $K$ is the so-called Rician Factor~\cite{3GPP38901, 3GPP36777}.
The \ac{LoS} component of the channel follows the plane-wave approximation~\cite{massivemimobook}, 
thus representing the phase shift of the plane wave with respect to each antenna element of the antenna panel.
It is computed as follows:
\begin{equation}\label{eq:ComplexChannelLOS}
    {\bf h}^{\rm LOS}_{u,b} = e^{
    - j \frac{2 \pi}{\lambda_c} d^{\rm 3D}_{u,b}
    }
    \; 
    e^{
    j \frac{2 \pi}{\lambda_c} \;
    {\bf k}_{u,b}^T\left( \phi_{u,b}, \theta_{u,b}  \right)
    \,
    {\bf U}_b
    }
\end{equation}
with 
\begin{equation}
    {\bf k}_{u,b} \in \mathbb{R}^{3 \times 1},\; {\bf U}_b \in \mathbb{R}^{3 \times M},
\end{equation}
where $\lambda_c$ is the frequency wavelength associated with the carrier frequency $f_c$, 
$d^{\rm 3D}_{u,b}$ is the 3D distance between \ac{UE} $u$ and the antenna panel centre of sector $b$, ${\bf k}_{u,b}^T\left( \phi_{u,b}, \theta_{u,b} \right)$ is the wave vector representing the plane wave variations in the 3D space, 
and ${\bf U}_b$ is the matrix containing the Cartesian coordinates of each antenna element w.r.t. the antenna panel centre.
The \ac{NLoS} component  of the channel is modeled as a Rayleigh fading complex channel as follows:
\begin{equation}\label{eq:ComplexChannelNLOS}
    {\bf h}^{\rm NLOS}_{u,b} \sim 
    \mathbb{CN}\left( {\bf 0}, {\bf I}_M \right)
    .
\end{equation}

\subsection{Cell Association and Precoding}\label{subsec:CellAssociation}

One of the physical layer features introduced in 5G \ac{NR} is the beamforming capability during the initial cell discovery phase via \ac{SSB} beams,
which allows sectors to cover different sections of their designated areas efficiently. 
Specifically, 
in a network operating in the sub-6 GHz band,
referred to as \ac{FR1}, 
each sector \(b\) can transmit up to 8 \ac{SSB} beams \cite{3GPP38214}. 
These beams are multiplexed sequentially in time, 
following a sweep pattern associated with their sweep index \(i_s^{\rm ssb}\).

\subsubsection*{SSB beams codebook}

At each sector \(b\), each \ac{SSB} beam \(s\) is represented by a complex codeword \({\bf w}_{s,b}^{\rm ssb}\) and is selected from a predetermined \ac{SSB} codebook \({\bf W}^{\rm ssb}\). 
We assume each antenna element of the planar array to be connected to a distinct transceiver. To accommodate beams with varying beamwidths and beamforming gains, 
we employ a switching pattern that sequentially deactivates antenna columns from the rightmost to the leftmost on the panel. 
For each configuration, 
we generate an intermediate \ac{SSB} codebook \({\bf W}^{\rm ssb}_i\) through a \ac{2D-DFT},
and subsequently aggregate these into the general \ac{SSB} codebook \({\bf W}^{\rm ssb}\), 
which consists of \(N_{\rm CB}\) codewords.

\subsubsection*{Cell association}

To identify its serving cell, 
each \ac{UE} $u$ measures the \ac{RSRP} from each cell $b$ and each \ac{SSB} beam $s$. 
The measured \ac{RSRP} is defined as follows:
\begin{equation}\label{eq:rsrpSSBComputation}
{\rm rsrp}^{\rm ssb}_{u, s, b} = \beta_{u,b} \; \left| {\bf h}^{\rm dl}_{u,b} \, {\bf w}_{s,b}^{\rm ssb} \right|^2 \; p_{s,b}^{\rm ssb} \; x_{s,b},  
\end{equation}
where $x_{s,b} \in {\bf X}$ is a binary variable, 
equal to one if beam $s$ is deployed at cell $b$,
and zero otherwise. 
The transmit power allocated by sector $b$ to beam $s$, denoted by $p_{s,b}^{\rm ssb}$, is an element of the matrix ${\bf P}$. 
The matrices ${\bf X}$ and ${\bf P}$ together provide a network-wide representation of the deployed beams and their transmit powers.
For each \ac{UE} $u$, 
the serving cell $\hat{b}_u$ and beam $\hat{s}_u$ are defined as those that maximize the measured \ac{RSRP}~\eqref{eq:rsrpSSBComputation}.

For each \ac{UE} $u$,
we define the coverage \ac{SINR} $\gamma^{\rm ssb}_u$ as follows:
\begin{equation} \label{eq:SINR_SSBCov}
    \gamma_u^{\rm ssb} = 
    \frac
    {{\rm rsrp}^{\rm ssb}_{u, \hat{s}_u, \hat{b}_u} }
    {
     \sum_{b=1, b \neq \hat{b}}^{N_{\rm BS}} \sum_{s=1}^{N_{\rm ssb}}
        {\rm rsrp}^{\rm ssb}_{u, s, b} \; \delta \left( i_{\hat{s}_u}^{\rm ssb} , i_s^{\rm ssb} \right) \; x_{s,b} 
    + N_u},
\end{equation}
where $\delta \left( i_{\hat{s}_u}^{\rm ssb} , i_s^{\rm ssb} \right)$ is defined as a binary function that takes a value of one if, 
and only if, $i_{\hat{s}_u}^{\rm ssb} = i_s^{\rm ssb}$.
\begin{figure*}[!htb] 
\normalsize
\setcounter{equation}{7}
\footnotesize
\begin{equation} 
 \label{eq:SINR_Computation}
\gamma_u = 
    \frac{
    \beta_{u, \hat{b}_u} \; \left| {\bf h}^{\rm dl}_{u, \hat{b}_u} \, {\bf w}^{\rm dl}_{u, \hat{b}_u} \right|^2   p^{\rm dl}_{u,\hat{b}_u}
    }
    {
    \beta_{u, \hat{b}_u}  \sum_{p \in \mathbb{U}_{\hat{b}_u} \setminus u}
    \left(1-\delta \left( {\bf w}^{\rm dl}_{u, \hat{b}_u} , {\bf w}^{\rm dl}_{p, \hat{b}_u} \right) \right)
     \left| {\bf h}^{\rm dl}_{u, \hat{b}_u} \, {\bf w}^{\rm dl}_{p, \hat{b}_u} \right|^2   p^{\rm dl}_{p,\hat{b}_u} + 
     \sum_{b \in B \setminus  \hat{b}_u} \beta_{u, b}
    \sum_{{\bf w}_{i,b}^{\rm dl} \in {\bf W}_b^{\rm dl}}
    \frac{1}{N_{{\bf w}^{\rm dl}_{i,b}}} \;\left| {\bf h}^{\rm dl}_{u, b} \, {\bf w}^{\rm dl}_{i, b}\right|^2  p^{\rm dl}_{i,b}  
    +   
    \frac{N_{\rm PRB} \; B_{\rm PRB}} {N_{{\bf w}^{\rm dl}_{u, \hat{b}_u}} } N_0
    }  
\end{equation}
\normalsize
\setcounter{equation}{8}
\hrulefill
\vspace*{4pt}
\end{figure*}

\subsection{SINR and Achievable Data Rate}

\subsubsection*{Data transmission phase}

To leverage the beamforming and multiplexing capabilities of \ac{NR} \ac{mMIMO} systems, 
we consider a Type I \ac{CSI}-based operational approach\mbox{\cite{DahlmanBook, 3GPP38214}}.
In this \ac{NR} network setup, 
each \ac{UE} reports a set of measurement indices to its serving cell. 
Based on these,
the sector chooses a specific codeword from a codebook, defined by \ac{2D-DFT} and considering all transceiver active, to precode the \ac{UE}'s data.
Specifically, 
the sector $b$ selects, for each \ac{UE} $u$, the downlink precoding vector ${\bf w}_{u, \hat{b}_u}^{\rm dl}$ as follows:
\begin{equation}
    {\bf w}_{u, \hat{b}_u}^{\rm dl} = 
    \argmax_{{\bf w} \in {\bf W}^{CB}}
    \left\{
        \beta_{u,b} \left| {\bf h}_{u, \hat{b}_u}^{\rm dl} \, {\bf w} \right|^2
    \right\}.
\end{equation}

\subsubsection*{SINR and achievable rate}

The resulting \ac{SINR} at \ac{UE} $u$ is computed according to \eqref{eq:SINR_Computation}.
In this framework, 
$\mathcal{U}_b$ represents the subset of \acp{UE} associated with cell $b$, 
and ${\bf h}_{u,b}^{\rm dl}$, ${\bf w}_{u,b}^{\rm dl}$, and $p_{u,b}^{\rm dl}$ are the downlink complex channel vector, the precoding codeword, and the associated transmit power of \ac{UE} $u$ with respect to cell $b$, respectively. Without loss of generality, we assume equal transmit power allocation for all \acp{UE}. 
The achievable data rate for each \ac{UE} can be then computed as follows:
\begin{equation}\label{eq:AchievableDataRate}
    R_u = \frac
    {N_{\rm PRB}^{\rm tot} \; B_{\rm PRB}}
    {N_{{\bf w}^{\rm dl}_{u, \hat{b}_u}}}
    \log_2(1 + \gamma_u),
\end{equation}
where 
$N_{{\bf w}^{\rm dl}_{u, \hat{b}_u}}$ is the number of \ac{UE} associated with the 
same precoding codeword and $N_0$ is the thermal noise power spectral density.

\section{Cell Selection and \ac{SSB} Beam Planning \label{sec:TwoStageAlgorithm}}

The performance of \ac{mMIMO} networks is affected by the complex interplay of many system parameters, making its modelling and large-scale optimization a challenging task.
To tackle this problem, we propose an efficient solution to maximize \acp{UAV} data rates along \ac{AH} by optimally controlling \acp{UAV} cell association along the \ac{AH}.
To this end, we first introduce a new metric to identify the optimal serving cell for each segment of an \ac{AH}. Then, we develop an \ac{eGA} algorithm to optimally select \ac{SSB} beams and their transmit power from a fixed codebook, ensuring optimal coverage from those cells.

\subsection{Aerial Highway Segment-to-Cell Association Metric}
In traditional cellular networks, 
serving cells are typically selected based on metrics such as \ac{RSRP}. 
While this metric may be suitable for \acp{gUE}, 
it often falls short for \acp{UAV} closely packed along the \ac{AH}, 
where the high channel correlation, driven by dominant \ac{LoS} conditions~\cite{GiuNikGer2024},
can severely affect network performance.
To optimally determine the cells designated to serve \acp{UAV} along the pre-defined \ac{AH}, 
we now introduce a novel metric that captures the multiplexing capability, average channel quality gain, and interference. 

\subsubsection*{Aerial highway Segmentation}

We begin by discretizing the \ac{AH} ${\bf r}_{\rm AH}$ into $N_r$ equidistant points separated by distance $d_r$. 
Utilizing simulations and/or measurements gathered during exploratory phases, one may determine the expected channel vector $\tilde{{\bf h}}_{r,b}$ for each point $r$ relative to each cell $b$ as follows, 
\begin{equation}\label{eq:expectedChannelVectorAH}
    \tilde{{\bf h}}_{r,b} =
    %
    \mathbb{E}_{\tau, {\bf h}^{\rm dl}} \,
    \left[ \,
    \rho_{r,b} \, \tau_{r,b} \, g_{r,b}  \, 
     {\bf h}^{\rm dl}_{r,b} 
    \, \right] .
\end{equation}
We utilize these vectors to construct the average complex channel matrix $\tilde{{\bf H}}_{r,m}^b \in \mathbb{C}^{N_r \times M}$ between the \ac{AH} and each cell $b$. 
Subsequently, we introduce the concept of a segment ${\bf z}$, 
which represents a contiguous subset of $N_s$ points within said \ac{AH} ${\bf r}_{\rm AH}$. 
Then, from matrix $\tilde{{\bf H}}_{r,m}^b$, we define two sub-matrices $\tilde{{\bf H}}_{z,m}^b$ and $\tilde{{\bf H}}_{r-z,m}^b$, respectively denoting the complex channel vectors of segment ${\bf z}$ and of the remaining \ac{AH} discrete points.

\subsubsection*{Cell association metric}

We define our proposed \ac{MAMA} metric as follows:
\begin{align}\label{eq:MetricCellAssociation}
    &\chi_{\bf z}^b 
    \left( \tilde{{\bf H}}_{z,m}^b, \tilde{{\bf H}}_{r-z,m}^b \right) = \\ \nonumber
    & \enspace= c_{\bf z}^b \left( \tilde{{\bf H}}_{z,m}^b \right) \;
    \log_2 \left(1+
    \frac
    {
    P_{\bf z}^b \left( \tilde{{\bf H}}_{z,m}^b \right)
    }
    {
    F_{\bf z}^b  \left( \tilde{{\bf H}}_{z,m}^b, \tilde{{\bf H}}_{r-z,m}^b \right)
    + N_0
    }
    \right) .
\end{align}
The metric in (\ref{eq:MetricCellAssociation}) is composed of three components, 
designed to account for different channel characteristics, specifically:
\begin{itemize}
    \item 
     $P_{\bf z}^b$ is the expected average channel gain on segment ${\bf z}$ when served by cell $b$. 
     It is computed as follows, 
        \begin{equation}\label{eq:MetricAverageChannelGain}
            P_{\bf z}^b  \left( \tilde{{\bf H}}_{z,m}^b \right) = 
            \frac{1}{N_z} \sum_z^{N_z} \;
            \frac{1}{M} \sum_{m=0}^{M-1} \left| h_{z,m}^b \right|^2,
        \end{equation}
    and it encapsulates traditional metrics like \ac{RSRP}.

    \item 
    $c_{\bf z}^b$ is the inverse of the condition number of matrix $\tilde{{\bf H}}_{z,m}^b$. 
    It is computed as follows:
    \begin{equation}\label{eq:MetricInvConditionNumber}
        c_{\bf z}^b \left( \tilde{{\bf H}}_{z,m}^b \right)  = 
        \frac{\lambda_{\bf z}^{b \, \left(M-1\right)}
        \left(\tilde{{\bf H}}_{z,m}^b \right)}
        {\lambda_{\bf z}^{b \, \left(0\right)} \left(\tilde{{\bf H}}_{z,m}^b \right)},
    \end{equation}
    where
    $\lambda_{\bf z}^{b \, \left(M-1\right)}$ and $\lambda_{\bf z}^{b \, \left(0\right)}$ 
    denote, respectively, the lowest and the highest singular values computed using \ac{SVD}. 
    This ratio provides insight into the spread of singular values, 
    reflecting diversity in \acp{AoA}/\acp{AoD} and assessing multiplexing capabilities of cell $b$ concerning segment ${\bf z}$.

    \item
    $F_{\bf z}^b$ represents the squared Frobenius norm of the cross-channel correlation. 
    It is computed as follows:
    \begin{align}\label{eq:MetricFrobNorm}
        &F_{\bf z}^b \left( \tilde{{\bf H}}_{z,m}^b,
        \tilde{{\bf H}}_{r-z,m}^b \right) =
        \sum_i^{N_r-N_z}  \sum_z^{N_z}
        \bigl| 
        \sum_m \tilde{h}_{i,m}^b \, \tilde{h}_{m,z}^{b\,*}
        \bigr|^2
        .
    \end{align}
    This component provides information about the correlation between the considered segment and the remaining points of the \ac{AH}, 
    thereby assessing the interference level introduced to other portions of the \ac{AH} when cell $b$ serves segment ${\bf z}$.
\end{itemize}

Having divided the \ac{AH} into $N_{\rm seg}$ segments, and having defined a cell association metric, one can then compute the serving cell $\hat{b}_{\bf z}$ for each segment ${\bf z}$ as follows:
\begin{equation}\label{eq:ServingCellForSegmentMetric}
    \hat{b}_{\bf z} = \argmax_{b \in \mathcal{B}} \left\{
    \chi_{\bf z}^b 
    \left( \tilde{{\bf H}}_{z,m}^b, \tilde{{\bf H}}_{r-z,m}^b \right) 
    \right\}.
\end{equation}
In the sequel, we denote by $\hat{{\bf b}}^{\rm (AH)}$ the set of serving cells for all segments.

\subsection{\ac{SSB} Beam and Power Selection Algorithm}
Given our proposed serving cell selection metric, 
we now propose an algorithm to optimally select the set of \ac{SSB} beams and their transmit power 
from the codebook ${\bf W}^{\rm ssb}$. These \ac{SSB} beams are then to be transmitted at each identified serving cell 
to guarantee the desired association of \acp{UAV} that are flying in segment ${\bf z}$. 
In other words,
our objective is to find the optimal binary entries of ${\bf X}$ and ${\bf P}$, governing the selected \acp{SSB} beams and their respective power,
that maximize a specific objective function ---later defined in \eqref{eq:Obj_Genetic}--- and, consequently, maximizing the minimum \ac{SINR}~\eqref{eq:SINR_SSBCov} across the \ac{AH}.

Before describing our solver, we find it convenient to define matrices ${\bf X}^{\rm bl}$ and ${\bf P}^{\rm bl}$, 
representing the network configuration in a traditional scenario with only \acp{gUE}. 
We also list the constraints that our algorithm obeys:
\begin{itemize}
    \item 
    Only serving cells in $\hat{{\bf b}}^{\rm (AH)}$ are permitted to modify their \ac{SSB} beam configuration. 
    Conversely, cells not in this set must adhere to the configurations described by ${\bf X}^{\rm bl}$ and ${\bf P}^{\rm bl}$.
    
    \item 
    Cells in $\hat{{\bf b}}^{\rm (AH)}$ are allowed to modify only one \ac{SSB} from the configuration in ${\bf X}^{\rm bl}$ and ${\bf P}^{\rm bl}$, 
    thus minimizing large deviations from the well optimized traditional scenario.
    
    \item 
    The power associated with each modified \ac{SSB} beam is limited to $p_{\rm max}^{\rm ssb}$.
    
    \item 
    The serving cell for each point $r$ of each segment ${\bf z}$ must be $\hat{b}_{\bf z}$.
\end{itemize}
To solve this problem, 
we design an algorithm based on \ac{eGA}~\cite{engelbrecht2007computational, GeneticAlgorithm, bhandari1996genetic}, 
known for its efficiency in solving non-convex non-linear mixed-integer problems.

\begin{algorithm}[!b]\label{algo:GeneticAlgorithm}
\caption{elite Genetic Algorithm Beam Selection}
\KwResult{${\bf y}_{\rm best}$}
${\bf y}^{(p)} \gets$ \small{init\_random\_population}$\left(N_{\rm pop},\, N_{\rm CB},\, p_{\rm max}^{\rm ssb} \right)$\; \vspace{2pt}
${\bf f} \gets {\rm init\_zeros}\left( N_{\rm pop}\right)$\;

\vspace{4pt}
\For{$i \in \left[0, N_{\rm eGA}^{\rm Iter}-1\right]$}{ \vspace{2pt}
    \For {$q \in \left[0, N_{\rm pop}-1\right]$}{
        ${\bf f}[q] \gets {\rm Obj}^{\rm eGA} \left({\bf y}^{(q)}\right)$\;
    }
    sort\_population$\left( {\bf f} \right)$\;
    ${\bf y}_{\rm best} \gets {\bf y}^{(0)}$\;

    ${\bf par}_e \gets \left[{\bf y}^{(0)}, {\bf y}^{(N_e)}    \right]$\; \vspace{2pt}

    ${\bf par}_q \gets \left[{\bf y}^{(0)}, {\bf y}^{(N_p)}    \right]$\;

    \For{$q \in N_{\rm cross}$}{
        ${\bf y}^{(i)}, {\bf y}^{(j)} \gets \text{\small{randomUniform\_selPair}} ({\bf par}_q)$\;

        \For{$k \in 0, \left[2n_s-1\right]$}{
            \If{${\rm random()}\leq P_{\rm cross} $}{
            ${\bf y}^{(i)}[k], {\bf y}^{(j)}[k] \gets {\bf y}^{(j)}[k], {\bf y}^{(i)}[k]$\;
            }
        } 
    }
    \For {$q \in 0, \left[N_{\rm pop}-1\right]$} {
        \For{$k \in \left[0, n_s-1\right]$}{
            \If{${\rm random()}\leq P_{\rm mut} $}{
            ${\bf y}^{(q)}[k] \gets \text{\small{randInt}}\left(0, N_{CB}-1 \right) $\;
            }
        }
        \For{$k \in \left[n_s, 2n_s-1\right]$}{
            \If{${\rm random()}\leq P_{\rm mut} $}{
            ${\bf y}^{(q)}[k] \gets \text{\small{rand}}\left(0, p_{\rm max}^{\rm ssb} \right) $\;
            }
        }
    }
    $e \gets 0$\;
    \For {$q \in  \left[N_{\rm pop}-N_e, N_{\rm pop}-1 \right]$}{
        ${\bf y}^{(q)} \gets {\bf par}_e [e]$\;
        $e \gets e + 1$\;
    }
    EarlyStopping\_Check$\left( i, N^{\rm stop}_{\rm eGA} \right)$\;
}
\end{algorithm} 

\subsubsection*{Preliminaries on eGA}
In \ac{eGA}, 
a solution emerges from a population of $N_{\rm pop}$ individuals, 
which iteratively transforms according to a objective (fitness) ${\rm Obj}^{\rm eGA}\left( \cdot \right)$ .
At each iteration, the population evolves following these steps:
\begin{itemize}
    \item 
    \textit{Selection}, 
    where individuals are ranked according to their fitness function, and $N_p$ best individuals are selected as parents for generating the offspring, i.e., next population individuals.
    \item 
    \textit{Crossover}, 
    where with a certain probability $P_{\rm cross}$ elements ${\bf y}_q$ of each pair of parents are randomly exchanged. We refer to these newly obtained vectors as offspring.
    \item 
    \textit{Mutation}, 
    where elements of the offspring vectors are randomly changed with probability $P_{\rm mut}$.
    \item 
    \textit{Elistic Mechanism}, 
    where the top $N_e$ individuals are directly passed to the next population without crossover and mutation mechanisms.
\end{itemize}
For each iteration, 
the optimal solution is selected as the individual with the best fitness value.
The algorithm progresses through these steps until it either reaches the designated number of generations, $N^{\rm Iter}_{\rm eGA}$, 
or the optimal solution remains unchanged over a specific number of generations
$N^{\rm stop}_{\rm eGA}$, therefore enabling an early stopping mechanism.

\subsubsection*{Proposed eGA beam selection algorithm}
For our problem, each individual $q$ of the population represents a possible solution and is defined by a vector as follows,
\begin{align}\label{eq:GeneticAlgoIndividualDefinition}
    &{\bf y}_q = 
    \left[     {\bf s}_{\hat{b}_{\bf z}}  \; \Big| \;    {\bf p}_{\hat{b}_{\bf z}}^{\rm ssb}     \right] = 
    \\ & \enspace=  \nonumber
    \left[ 
            s_{b_0}, \ldots, s_{b_{\bf z}}, \ldots, s_{b_{n_s}} \; \Big| \; p_{b_0}^{\rm ssb}, \ldots, p_{b_{\bf z}}^{\rm ssb}, \ldots, p_{b_{n_s}}^{\rm ssb}
    \right], 
\end{align}
where $s_{b_{\bf z}} \in \left[ 0, N_{\rm CB} \right]$ represents the codeword index selected at cell $b_{\bf z}$ from a codebook containing $N_{\rm CB}$ entries, 
and $ p_{b_{\bf z}}^{\rm ssb}$ represents its transmit power.
The matrices ${\bf X}_q$ and ${\bf P}_q$ are then computed 
based on the above vector values.
To evaluate the performance of each individual $q$, 
we define the objective (fitness) function as
\begin{equation}
\label{eq:Obj_Genetic}
     {\rm Obj}^{\rm eGA}\left( {\bf y}_q \right) \doteq 
    {\rm min}\left\{
       \gamma_z^{\rm ssb} \,
        \big| \,
        z \in {\bf z} \,,
        \, \hat{b}_z = \hat{b}_{\bf z}^{\rm (AH)},
        {\bf X}_q, {\bf P}_q 
    \right\} .
\end{equation}
The objective function in~\eqref{eq:Obj_Genetic} represents the minimum \ac{SINR} across all the \ac{AH} to be maximized. 
In particular, 
$\gamma_z^{\rm ssb}$ is the \ac{SSB} \ac{SINR} for point $z$, 
as defined in~\eqref{eq:SINR_SSBCov}, conditioned on $\hat{b}_{\bf z}^{\rm (AH)}$ being the serving cell for point $z$ 
and ${\bf X}_q$ and ${\bf P}_q$ being the selected \acp{SSB} beams and their transmit powers, respectively.

Algorithm~\ref{algo:GeneticAlgorithm} illustrates the detailed steps.

\section{Simulation Results}\label{sec:Results}

In this section, 
we evaluate the performance of our proposed \ac{SSB} beam planning and UAV cell
selection.
Utilizing the models presented in Section~\ref{sec:SystemModel}, 
we focus on \ac{UMa} environments where each sector employs an $8 \times 4$ \ac{UPA} panel operating at $3.5$\,GHz. 
The study considers a 1250\,m \ac{AH} located 100\,m above the ground and crossing multiple cell edges. 
Then, ensuring 100\,m inter-\acp{UAV} distance $d_{\rm IUD}$~\cite{10008607},
12 \acp{UAV} are evenly spaced along this highway, 
and four \acp{gUE} are randomly deployed within each cell.

\subsubsection*{Performance benchmark}

We compare the results obtained by our solution (``Opt'') to those of a baseline configuration 
where the terrestrial network is optimized solely for serving \acp{gUE} (``Baseline''). 
The baseline configuration positions all \ac{SSB} beams at each cell with a tilt of $105^\circ$, covering the azimuth plane as recommended by the \ac{3GPP}~\cite{3GPP38901}. The matrices ${\bf X}^{\rm bl}$ and ${\bf X}^{\rm bl}$ are then computed accordingly.

\subsubsection*{Algorithm parameters and convergence}

In Algorithm~\ref{algo:GeneticAlgorithm}, we choose a population size $N_{\rm pop} = 100$, a number of parents $N_p = 75$, a number of elites $N_e = 20$, and we set the probabilities of crossover and mutation to $P_{\rm cross}=0.2$ and $P_{\rm mut} = 0.75$, respectively.
We set the maximum number of iterations to $N^{\rm Iter}_{\rm eGA}=15000$ and the early stopping criterion to $N^{\rm stop}_{\rm eGA}=1000$ iterations. Under these settings, our algorithm converges after 12000 iterations.

\subsubsection*{SINR and achievable data rate}

\begin{figure*}[!ht]
    \centering
    \subfloat [][\footnotesize{\ac{SINR} Distribution}\label{subfig:SINR_12UMaSingleEdge100}]{
        \resizebox{0.48\textwidth}{!}{\input{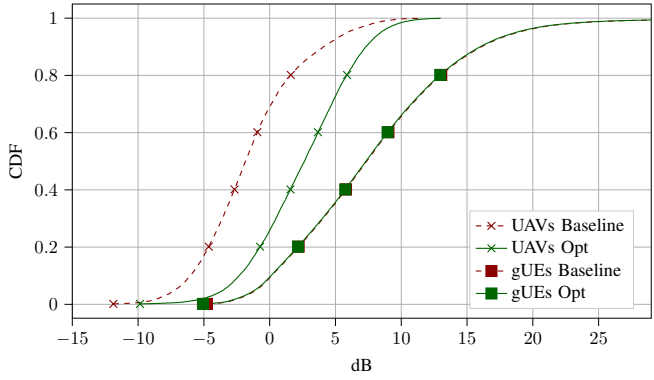}}
    }
    \hfill 
    \subfloat[][\footnotesize{Achievable Data Rate Distribution}\label{subfig:RATE_12UMaSingleEdge100}]{
        \resizebox{0.48\textwidth}{!}{\input{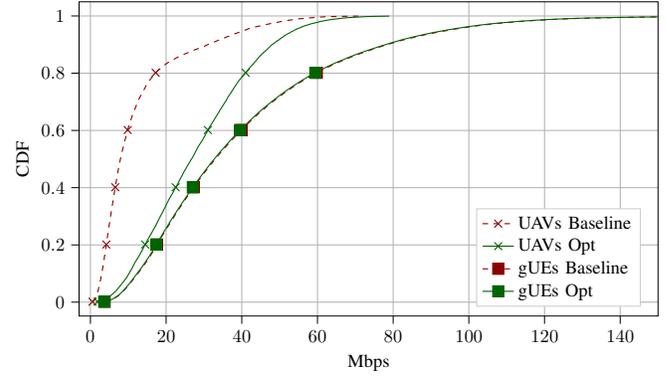}}
    }
    \caption{\ac{SINR} and achievable data rate distribution for 12 \acp{UAV} on an \ac{AH} positioned across cell edges at a height of 100\,m. 
    \label{fig:SINR_RATE_12UMaSingleEdge100}}
\vspace{-0.75em}
\end{figure*}

Figure~\ref{fig:SINR_RATE_12UMaSingleEdge100} displays the \ac{CDF} of (a) \ac{SINR} and (b) achievable data rates, distinguishing between \acp{UAV} and \acp{gUE}. 
The results show an improvement of $3.7$\,dB in the 5\%-tile \ac{SINR} for \acp{UAV}, 
moving from $-7.21$\,dB (Baseline) to $-3.51$\,dB (Opt). 
Moreover, our solution offers a four-fold increase in the $5$\%-tile achievable data rate for \acp{UAV}, 
rising from $2$\,Mbps (Baseline) to $8$\,Mbps (Opt). 
Similar gains are observed in the mean \ac{SINR} and mean achievable data rate.
Furthermore, 
by varying only a single \ac{SSB} beam from the baseline configuration
while optimizing the network for \acp{UAV}, 
we incur a very limited \acp{gUE} performance degradation of 0.15\,dB in the 5\%-tile \ac{SINR} and 1\,\% in the the 5\%-tile achievable data rate. 

\subsubsection*{Traffic analysis}

\begin{figure}[!ht]
    \centering
    \subfloat{
        \resizebox{0.47\textwidth}{!}{
\begin{tikzpicture}

\definecolor{darkgray176}{RGB}{176,176,176}
\definecolor{darkred14080}{RGB}{140,8,0}
\definecolor{forestgreen1811328}{RGB}{18,113,28}
\definecolor{lightgray204}{RGB}{204,204,204}
\definecolor{navy028127}{RGB}{0,28,127}

\begin{axis}[
width=5in,
height=3in,
legend cell align={left},
legend style={fill opacity=0.8, draw opacity=1, text opacity=1, draw=lightgray204},
tick align=outside,
tick pos=left,
x grid style={darkgray176, opacity=0.45},
xlabel={Inter-UAV distance \(\displaystyle d_\mathrm{IUD}\) [m]},
xmajorgrids,
xmin=-1.45, xmax=52.45,
xtick style={color=black},
xtick={1,4,7,10,13,16,19,22,25,28,31,34,37,40,43,46,49},
xticklabel style={rotate=45.0},
xticklabels={
  1250.0,
  312.5,
  178.57,
  125.0,
  96.15,
  78.12,
  65.79,
  56.82,
  50.0,
  44.64,
  40.32,
  36.76,
  33.78,
  31.25,
  29.07,
  27.17,
  25.51
},
y grid style={darkgray176, opacity=0.45},
ylabel={Mbps},
ymajorgrids,
ymin=-0.62074918816378, ymax=15,
ytick style={color=black}
]
\addplot [semithick, darkred14080, opacity=1, dashed, mark=triangle*, mark size=3, mark repeat=3, mark options={solid,rotate=180}]
table {%
1 9.15557896345051
2 8.99005037097571
3 7.31450331684988
4 5.94685026215204
5 5.05243103742135
6 4.4919466301346
7 3.95455623042464
8 3.37882696432563
9 2.8967933650492
10 2.68811885300129
11 2.41201541104729
12 2.19583638827521
13 2.07425021347427
14 1.85486388736323
15 1.71141387184841
16 1.56984061743
17 1.45601852592188
18 1.35104755015489
19 1.23097207582454
20 1.13683468002485
21 1.06781598181067
22 0.967415186455231
23 0.911406147458748
24 0.863478720548572
25 0.796742831238184
26 0.750020636930568
27 0.70594210171663
28 0.674345884156513
29 0.637121614749731
30 0.606819194746276
31 0.575092450622012
32 0.535482962704567
33 0.526106519961509
34 0.50103858448315
35 0.476659836079812
36 0.459950693341482
37 0.441662932145992
38 0.425372598703333
39 0.412758094356486
40 0.398840045914744
41 0.380108904052702
42 0.368399160722339
43 0.363170069270386
44 0.353571661876216
45 0.331534711768194
46 0.325454301056021
47 0.314041454461347
48 0.303352528857283
49 0.296485539441343
50 0.294083006880079
};
\addlegendentry{UAV Baseline}

\addplot [semithick, forestgreen1811328, opacity=1, mark=x, mark size=3, mark repeat=3, mark options={solid}]
table {%
1 14.3985840065796
2 14.0291420550444
3 13.8167843434765
4 12.8153035344138
5 11.4980417452696
6 11.118441051173
7 10.8943815682919
8 10.4071253485783
9 9.52434824881337
10 8.83832238792517
11 8.27799264061878
12 7.45124239239064
13 7.00655977317431
14 6.22657338874207
15 5.53379476656637
16 4.73431119414176
17 4.13574469862041
18 3.77932863421973
19 3.60013104099327
20 3.3634117502404
21 3.17907214801094
22 3.00834625494196
23 2.93334299994231
24 2.75255336394761
25 2.56773123804506
26 2.43500639003026
27 2.3637602560271
28 2.18640443468806
29 2.1360266546359
30 2.06162304627327
31 1.91684436593028
32 1.88620746763693
33 1.79223465667236
34 1.73226158590041
35 1.6545162542358
36 1.65208732912516
37 1.56591892427688
38 1.50939537647917
39 1.48917371357895
40 1.41013417031378
41 1.39726900447447
42 1.34376673929281
43 1.31343070094889
44 1.2757142844559
45 1.22095150388055
46 1.19721613920091
47 1.1562034938122
48 1.10354174252512
49 1.08178860561445
50 1.07022121956161
};
\addlegendentry{UAV Opt}
\end{axis}

\begin{axis}[
width=5in,
height=3in,
axis x line=top,
tick align=outside,
x grid style={darkgray176, opacity=0.45},
xlabel={Number of UAVs},
xmin=-1.45, xmax=52.45,
xtick pos=right,
xtick style={color=black},
xtick={1,4,7,10,13,16,19,22,25,28,31,34,37,40,43,46,49},
y grid style={darkgray176, opacity=0.45},
ymin=-0.62074918816378, ymax=15,
ytick pos=left,
ytick style={color=black}
]
\addplot [semithick, darkred14080, opacity=1, dashed, mark=triangle*, mark size=3, mark repeat=3, mark options={solid,rotate=180}]
table {%
1 9.15557896345051
2 8.99005037097571
3 7.31450331684988
4 5.94685026215204
5 5.05243103742135
6 4.4919466301346
7 3.95455623042464
8 3.37882696432563
9 2.8967933650492
10 2.68811885300129
11 2.41201541104729
12 2.19583638827521
13 2.07425021347427
14 1.85486388736323
15 1.71141387184841
16 1.56984061743
17 1.45601852592188
18 1.35104755015489
19 1.23097207582454
20 1.13683468002485
21 1.06781598181067
22 0.967415186455231
23 0.911406147458748
24 0.863478720548572
25 0.796742831238184
26 0.750020636930568
27 0.70594210171663
28 0.674345884156513
29 0.637121614749731
30 0.606819194746276
31 0.575092450622012
32 0.535482962704567
33 0.526106519961509
34 0.50103858448315
35 0.476659836079812
36 0.459950693341482
37 0.441662932145992
38 0.425372598703333
39 0.412758094356486
40 0.398840045914744
41 0.380108904052702
42 0.368399160722339
43 0.363170069270386
44 0.353571661876216
45 0.331534711768194
46 0.325454301056021
47 0.314041454461347
48 0.303352528857283
49 0.296485539441343
50 0.294083006880079
};

\addplot [semithick, forestgreen1811328, opacity=1, mark=x, mark size=3, mark repeat=3, mark options={solid}]
table {%
1 14.3985840065796
2 14.0291420550444
3 13.8167843434765
4 12.8153035344138
5 11.4980417452696
6 11.118441051173
7 10.8943815682919
8 10.4071253485783
9 9.52434824881337
10 8.83832238792517
11 8.27799264061878
12 7.45124239239064
13 7.00655977317431
14 6.22657338874207
15 5.53379476656637
16 4.73431119414176
17 4.13574469862041
18 3.77932863421973
19 3.60013104099327
20 3.3634117502404
21 3.17907214801094
22 3.00834625494196
23 2.93334299994231
24 2.75255336394761
25 2.56773123804506
26 2.43500639003026
27 2.3637602560271
28 2.18640443468806
29 2.1360266546359
30 2.06162304627327
31 1.91684436593028
32 1.88620746763693
33 1.79223465667236
34 1.73226158590041
35 1.6545162542358
36 1.65208732912516
37 1.56591892427688
38 1.50939537647917
39 1.48917371357895
40 1.41013417031378
41 1.39726900447447
42 1.34376673929281
43 1.31343070094889
44 1.2757142844559
45 1.22095150388055
46 1.19721613920091
47 1.1562034938122
48 1.10354174252512
49 1.08178860561445
50 1.07022121956161
};
\end{axis}

\end{tikzpicture}}
    }
\caption{ 5\%-tile achievable data rate for different traffic density and $d_{\rm IUD}$. 
}

\label{fig:SINR_Rate_TrafficAnalysis}
\vspace{-1.5em}
\end{figure}
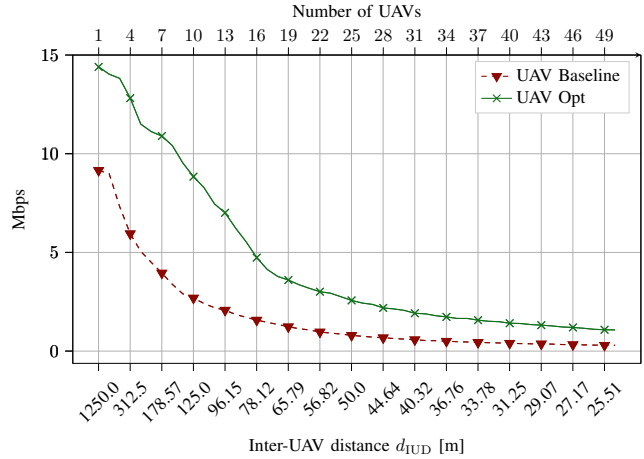

Figure~\ref{fig:SINR_Rate_TrafficAnalysis} illustrates how the  5\%-tile achievable data rate evolves as the traffic on the \ac{AH} increases. 
For this analysis, 
we consider an increasing number of flying \acp{UAV}, up to 50, evenly spaced along the \ac{AH}. 
When imposing a minimum data rate threshold of 5\,Mbps, 
the baseline configuration can support only up to 5 \acp{UAV}. 
In contrast, our optimized solution can accommodate up to 15 \acp{UAV}, 
effectively tripling the traffic capacity on the \ac{AH}. 

\section{Conclusion}\label{sec:Conclusion}

In this paper, 
we investigated how to provide optimal 5G connectivity from terrestrial \ac{mMIMO} networks to \acp{UAV} within \acp{AH}, 
while minimizing impact on ground performance.
We proposed a metric that optimally determines the serving cells for multiple segments of the \acp{AH} by jointly considering multiplexing capabilities, channel gains, and interference. 
Following this, 
we developed an algorithm to identify the optimal \ac{SSB} beam planning strategy, 
thereby ensuring optimal coverage of the \acp{AH} from the selected cells.
Simulation results demonstrated the benefits of our approach, with gains of up to four times in achievable data rates for \acp{UAV}. 
This illustrates that strategically and optimally controlling the selection of serving cells along the \ac{AH} is crucial for enhancing \acp{UAV} capacity with minimal impact on ground performance. 
While in this paper we optimized the \acp{SSB} beams planning given a fixed \ac{AH} partition, further performance gains may be achieved by optimizing the \ac{AH} segmentation as well.

\begin{acronym}[AAAAAAAAA]
    \acro{CSI-RS} {channel state information-reference signal}
    \acro{SRS} {sounding reference signal}    
    \acro{P2P} {point to point}    
    \acro{UPA} {uniform planar array}
    \acro{PMI} {precoding matrix indicator}
    \acro{RI} {rank indicator}
    \acro{CQI} {channel quality indicator}
    \acro{DL} {downlink}
    \acro{CDF}{cumulative distribution function}
    \acro{DFT}{discrete Fourier transform}
    \acro{2D-DFT}{two dimensional discrete Fourier transform}
    \acro{NR}{new radio}
    \acro{PL}{path loss}
    \acro{AF}{array factor}
    \acro{AH}{aerial highway}
    \acro{PSO}{particle swarm optimization}
    \acro{eGA}{elite genetic algorithm}
    \acro{GA}{genetic algorithm}
    \acro{ICC}{international conference on communications}
    \acro{SSB}{synchronization signal block}
    \acro{ES}{Eigenscore}
    \acro{AoA}{angle of arrival}
    \acro{AoD}{angle of departure}
    \acro{UAV}{unmanned aerial vehicle}
    \acro{CCUAV}{cellular connected unmanned aerial vehicle}
    \acro{D2D}{device to device}
    \acro{gUE}{ground user equipment}
    \acro{UE}{user equipment}
    \acro{MIMO}{multiple-input multiple-output}
    \acro{mMIMO}{massive multiple-input multiple-output}
    \acro{MU-mMIMO}{multi-user massive multiple-input multiple-output}
    \acro{MU-MIMO}{multi-user multiple-input multiple-output}
    \acro{SU-MIMO}{single-user multiple-input multiple-output}
    \acro{PRB}{physical resource block}
    \acro{RE}{resource element}
    \acro{RSRP}{reference signal received power}
    \acro{RSS}{received signal strength}
    \acro{mmWave}{millimetre wave}
    \acro{eICIC}{enhanced inter-cell interference coordination}
    \acro{SINR}{signal-to-interference-plus-noise ratio}
    \acro{UAM}{urban air mobility}
    \acro{QoS}{quality of services}
    \acro{UE}{user equipment}
    \acro{LoS}{line of sight}
    \acro{NLoS}{non-line of sight}
    \acro{BVLoS}{beyond visual line of sight}
    \acro{DoF}{degree of freedom}
    \acro{ZF}{zero forcing}
    \acro{CSI}{channel state information}
    \acro{3GPP}{3rd Generation Partnership Project}
    \acro{SVD}{single value decomposition}
    \acro{PBCH}{physical broadcast channel}
    \acro{thp}{throughput}
    \acro{UMi}{urban micro}
    \acro{UMa}{urban macro}
    \acro{CAGR}{compound annual growth rate}
    \acro{HO}{handover}
    \acro{MNO}{mobile network operator}
    \acro{NOMA}{non-orthogonal multiple access}
    \acro{BO}{bayesan optimization}
    \acro{ML}{machine learning}
    \acro{FR1}{frequency range 1}
    \acro{SO}{southern}
    \acro{E}{eastern}
    \acro{NE}{northeastern}
    \acro{RAN}{radio access network}
    \acro{BS}{base station}
    \acro{ISD}{inter-site distance}
    \acro{IUD}{inter-UAV distance}
    \acro{RRC}{radio resource control}
    \acro{PSS}{primary synchronization signal}
    \acro{SSS}{secondary synchronization signal}
    \acro{PBCH}{physical broadcasted channel}
    \acro{MINP}{mixed-integer nonlinear problem}
    \acro{PAHSS}{Particle Aerial Highway Swarm Segmentation}
    \acro{MAMA}{mMIMO-Aerial-Metric-Association}
    \acro{URD}{Urban Random Distributed}
    \acro{UDN}{ultra dense network}
\end{acronym}

\bibliographystyle{IEEEtran}
\bibliography{journalAbbreviations, bibl}

\begin{thebibliography}{10}
\providecommand{\url}[1]{#1}
\csname url@samestyle\endcsname
\providecommand{\newblock}{\relax}
\providecommand{\bibinfo}[2]{#2}
\providecommand{\BIBentrySTDinterwordspacing}{\spaceskip=0pt\relax}
\providecommand{\BIBentryALTinterwordstretchfactor}{4}
\providecommand{\BIBentryALTinterwordspacing}{\spaceskip=\fontdimen2\font plus
\BIBentryALTinterwordstretchfactor\fontdimen3\font minus \fontdimen4\font\relax}
\providecommand{\BIBforeignlanguage}[2]{{%
\expandafter\ifx\csname l@#1\endcsname\relax
\typeout{** WARNING: IEEEtran.bst: No hyphenation pattern has been}%
\typeout{** loaded for the language `#1'. Using the pattern for}%
\typeout{** the default language instead.}%
\else
\language=\csname l@#1\endcsname
\fi
#2}}
\providecommand{\BIBdecl}{\relax}
\BIBdecl

\bibitem{zeng2020uav}
Y.~Zeng, I.~Guvenc, R.~Zhang, G.~Geraci, and D.~W. Matolak, \emph{UAV Communications for 5G and Beyond}.\hskip 1em plus 0.5em minus 0.4em\relax John Wiley \& Sons, 2020.

\bibitem{Wu2021}
Q.~Wu, J.~Xu, Y.~Zeng, D.~W.~K. Ng, N.~Al-Dhahir, R.~Schober, and A.~L. Swindlehurst, ``A comprehensive overview on {5G}-and-beyond networks with {UAVs}: From communications to sensing and intelligence,'' \emph{{IEEE} J. Sel. Areas Commun.}, vol.~39, no.~10, pp. 2912--2945, 2021.

\bibitem{8470897}
Y.~Zeng, J.~Lyu, and R.~Zhang, ``Cellular-connected {UAV}: Potential, challenges, and promising technologies,'' \emph{{IEEE} Wireless Commun.}, vol.~26, no.~1, pp. 120--127, Sep. 2019.

\bibitem{KanMezLoz2021}
S.~Kang, M.~Mezzavilla, A.~Lozano, G.~Geraci, W.~Xia, S.~Rangan, V.~Semkin, and G.~Loianno, ``Millimeter-wave {UAV} coverage in urban environments,'' in \emph{Proc. IEEE Globecom}, 2021.

\bibitem{MarketGrowthReport}
{Market Growth Reports}, ``{Global Civil Drone Industry Research Report 2023, Competitive Landscape, Market Size, Regional Status and Prospect},'' January 2023.

\bibitem{PilotsHandbook_FAA}
{F.A.A.}, \emph{{Pilot's Handbook of Aeronautical knowledge}}.\hskip 1em plus 0.5em minus 0.4em\relax U.S. Department of Transportation, 2016.

\bibitem{9566514}
N.~Cherif, W.~Jaafar, H.~Yanikomeroglu, and A.~Yongacoglu, ``{3D} aerial highway: The key enabler of the retail industry transformation,'' \emph{{IEEE} Commun. Mag.}, vol.~59, no.~9, pp. 65--71, Sep. 2021.

\bibitem{9768113}
G.~Geraci, A.~Garcia-Rodriguez, M.~M. Azari, A.~Lozano, M.~Mezzavilla, S.~Chatzinotas, Y.~Chen, S.~Rangan, and M.~D. Renzo, ``What will the future of {UAV} cellular communications be? {A} flight from {5G} to {6G},'' \emph{{IEEE} Commun. Surveys Tuts.}, vol.~24, no.~3, pp. 1304--1335, 2022.

\bibitem{9681624}
M.~Mozaffari, X.~Lin, and S.~Hayes, ``Toward {6G} with connected sky: {UAVs} and beyond,'' \emph{{IEEE} Commun. Mag.}, vol.~59, no.~12, pp. 74--80, 2021.

\bibitem{geraci2022integrating}
G.~Geraci, D.~L{\'o}pez-P{\'e}rez, M.~Benzaghta, and S.~Chatzinotas, ``{Integrating terrestrial and non-terrestrial networks: {3D} opportunities and challenges},'' \emph{{IEEE} Commun. Mag.}, vol.~61, no.~4, pp. 42--48, 2023.

\bibitem{9839125}
A.~Colpaert, M.~Raes, E.~Vinogradov, and S.~Pollin, ``Drone delivery: Reliable cellular {UAV} communication using multi-operator diversity,'' in \emph{Proc. IEEE Int. Conf. on Comm. (ICC)}, Aug. 2022, pp. 1--6.

\bibitem{DoPulFod2024}
H.~Do, R.~D. Pulgar, G.~Fodor, and Z.~Qi, ``Cellular connectivity for advanced air mobility: Use cases and beamforming approaches,'' \emph{{IEEE} Commun. Std. Mag.}, vol.~8, no.~1, pp. 65--71, 2024.

\bibitem{10021675}
S.~J. Maeng, M.~M.~U. Chowdhury, I.~Güvenç, A.~Bhuyan, and H.~Dai, ``Base station antenna uptilt optimization for cellular-connected drone corridors,'' \emph{IEEE Trans. Aerospace Elec. Systems}, vol.~59, no.~4, pp. 4729--4737, Jan. 2023.

\bibitem{chowdhury2021ensuring}
M.~M.~U. Chowdhury, I.~Guvenc, W.~Saad, and A.~Bhuyan, ``Ensuring reliable connectivity to cellular-connected {UAVs} with up-tilted antennas and interference coordination,'' \emph{ITU J. Fut. and Evol. Technol.}, pp. 165--185, Dec. 2021.

\bibitem{10001469}
M.~Bernabè, D.~Lopez-Perez, D.~Gesbert, and H.~Bao, ``On the optimization of cellular networks for {UAV} aerial corridor support,'' in \emph{Proc. IEEE Global Commun. Conf. (GLOBECOM)}, Jan. 2022, pp. 2969--2974.

\bibitem{karimibidhendi2023optimizing}
S.~Karimi-Bidhendi, G.~Geraci, and H.~Jafarkhani, ``Optimizing cellular networks for {UAV} corridors via quantization theory,'' \emph{arXiv preprint arXiv:2308.01440}, 2023.

\bibitem{benzaghta2023designing}
M.~Benzaghta, G.~Geraci, D.~López-Pérez, and A.~Valcarce, ``Designing cellular networks for {UAV} corridors via {Bayesian} optimization,'' in \emph{Proc. IEEE Global Commun. Conf. (GLOBECOM)}, Feb. 2023, pp. 4552--4557.

\bibitem{8528463}
G.~Geraci, A.~Garcia-Rodriguez, L.~Galati~Giordano, D.~López-Pérez, and E.~Björnson, ``Understanding {UAV} cellular communications: From existing networks to massive {MIMO},'' \emph{IEEE Access}, vol.~6, Nov. 2018.

\bibitem{8869706}
A.~Garcia-Rodriguez, G.~Geraci, D.~Lopez-Perez, L.~G. Giordano, M.~Ding, and E.~Bjornson, ``The essential guide to realizing {5G}-connected {UAVs} with massive {MIMO},'' \emph{{IEEE} Commun. Mag.}, vol.~57, no.~12, pp. 84--90, Oct. 2019.

\bibitem{9374639}
Y.~Huang, Q.~Wu, R.~Lu, X.~Peng, and R.~Zhang, ``Massive {MIMO} for cellular-connected {UAV}: Challenges and promising solutions,'' \emph{{IEEE} Commun. Mag.}, vol.~59, no.~2, pp. 84--90, Feb. 2021.

\bibitem{7314981}
D.~Bethanabhotla, O.~Y. Bursalioglu, H.~C. Papadopoulos, and G.~Caire, ``Optimal user-cell association for massive {MIMO} wireless networks,'' \emph{IEEE Trans. Wireless Commun.}, vol.~15, no.~3, pp. 1835--1850, Nov. 2016.

\bibitem{7247514}
A.~G. Gotsis, S.~Stefanatos, and A.~Alexiou, ``Optimal user association for massive {MIMO} empowered ultra-dense wireless networks,'' in \emph{Proc. IEEE Int. Conf. on Comm. Workshops (ICC Workshops)}, 2015, pp. 2238--2244.

\bibitem{10569086}
M.~Bernab{è}, D.~L{ó}pez-P{é}rez, N.~Piovesan, G.~Geraci, and D.~Gesbert, ``Massive {MIMO} for aerial highways: Enhancing cell selection via {SSB} beams optimization,'' \emph{{IEEE} O.J. on Commun.}, pp. 1--1, Jun. 2024.

\bibitem{3GPP38901}
{3GPP TR38.901}, ``{Study on channel model for frequencies from 0.5 to 100 GHz},'' Mar. 2017, v.14.0.

\bibitem{3GPP36777}
{3GPP TR36.777}, ``{Enhanced LTE support for aerial vehicles},'' Jan. 2017, v.15.0.

\bibitem{massivemimobook}
E.~Björnson, J.~Hoydis, and L.~Sanguinetti, \emph{Massive {MIMO} Networks: Spectral, Energy, and Hardware Efficiency}.\hskip 1em plus 0.5em minus 0.4em\relax Now Foundations and Trends, 2017.

\bibitem{3GPP38214}
{3GPP TR38.214}, ``{Phisical layer procedure for data},'' Sep. 2023, v.18.0.

\bibitem{DahlmanBook}
E.~Dahlman, S.~Parkvall, and J.~Skold, \emph{{5G} {NR}: The Next Generation Wireless Access Technology}, 2nd~ed.\hskip 1em plus 0.5em minus 0.4em\relax USA: Academic Press, Inc., 2018.

\bibitem{GiuNikGer2024}
A.~Giuliani, R.~Nikbakht, G.~Geraci, S.~Kang, A.~Lozano, and S.~Rangan, ``Spatially consistent air-to-ground channel modeling via generative neural networks,'' \emph{{IEEE} Commun. Lett.}, vol.~13, no.~4, pp. 1158--1162, 2024.

\bibitem{engelbrecht2007computational}
A.~P. Engelbrecht, \emph{Computational intelligence: an introduction}.\hskip 1em plus 0.5em minus 0.4em\relax John Wiley \& Sons, 2007.

\bibitem{GeneticAlgorithm}
J.~H. Holland, ``Genetic algorithms,'' \emph{Scientific American}, vol. 267, no.~1, pp. 66--73, 1992.

\bibitem{bhandari1996genetic}
D.~Bhandari, C.~Murthy, and S.~K. Pal, ``Genetic algorithm with elitist model and its convergence,'' \emph{Internat. Journal of Pattern Recogn. and A.I.}, vol.~10, no.~06, pp. 731--747, 1996.

\bibitem{10008607}
E.~Vinogradov and S.~Pollin, ``Reducing safe {UAV} separation distances with {U2U} communication and new remote {ID} formats,'' in \emph{Proc. IEEE Global Commun. Conf. Workshops (GLOBECOM Workshops)}, Jan. 2022, pp. 1425--1430.

\end{thebibliography}


\end{document}